\documentclass[12pt,a4paper]{article}
\usepackage{amsfonts,latexsym}
\usepackage{graphicx,color}
\usepackage{dcolumn}
\usepackage{graphicx}
\usepackage{amsmath}
\usepackage{amsfonts}
\usepackage{amssymb}

\renewcommand{\thefootnote}{\fnsymbol{footnote}}

\begin{document}

\vspace{12mm}

\begin{center}
{{{\Large {\bf Scalarized Einstein-Euler-Heisenberg black holes at the approach to extremality}}}}\\[10mm]
{Hong Guo$^1$\footnote{e-mail address: hong\_guo@pku.edu.cn}, Xiao Yan Chew$^2$\footnote{e-mail address: xiao.yan.chew@just.edu.cn}, and Yun Soo Myung$^{3}$\footnote{e-mail address: ysmyung@inje.ac.kr}}\\[8mm]
{${}^1$Kavli Institute for Astronomy and Astrophysics, Peking University, Beijing 100871, China\\[0pt]}
{${}^2$School of Science, Jiangsu University of Science and Technology, Zhenjiang, 212100, China\\[0pt]}
{${}^3${Center for Quantum Spacetime, Sogang University, Seoul 04107, Republic of  Korea\\[0pt] }}
\end{center}

\vspace{1mm}

\begin{abstract}
We investigate extremal black holes with scalar hair in the generic Einstein-Euler-Heisenberg (EEH)-scalar theory with two scalar couplings to the Maxwell term. 
One is an exponential coupling with coupling constant  $\alpha$ and the other is its polynomial coupling.   
We first construct scalarized black holes on the cold (C)-horizon of EEH black holes described by mass $M$ and magnetic charge $P$ both at the linear level, through the existence curves $\alpha_n(q)$ with $q=P/M$, and as fully backreaction solutions.
As extremality ($q=q_e$) is approached, all existence curves accumulate, following $\alpha_n-\alpha_c\sim 1/\ln^2(q_e-q)$, at  the critical branch  $\alpha_c$ fixed by the Breitenlohner-Freedman bound for the near-horizon (AdS$_2\times S^2$) throat.
We obtain  scalarized extremal black hole (SEBH) with constant secondary hair and it is recovered exactly from the entropy function approach working on the near-horizon throat.
Its entropy is an attractor invariant, being independent of the asymptotic modulus, so the scalar  hair remains secondary.
Exploiting this constant scalar, we seek  further SEBHs with  scalar hair keeping  its charge $Q_s$  for the zero asymptotic scalar ($\phi_\infty=0$).
In the $(q,\alpha)$ plane, we observe that these extremal solutions occupy $q\ge q_e$. Hence, the existence curves $\alpha_n(q)$ existing for $q\le q_e$ and the extremal branches  bound the scalarized domain from opposite sides and they meet only at $q=q_e$.
\end{abstract}
\vspace{4mm}

\hspace{11.5cm}
\newpage
\renewcommand{\thefootnote}{\arabic{footnote}}
\setcounter{footnote}{0}


\section{Introduction}
 A nonminimal scalar coupling to the Gauss-Bonnet (GB) term  has induced  spontaneous scalarization of Schwarzschild black hole triggered by tachyonic scalar~\cite{Doneva:2017bvd,Silva:2017uqg,Antoniou:2017acq}.
 This approach  has usually  provided   a way to obtain   black holes with secondary scalar hair  as well as  it indicated  a mechanism to understand the interaction between gravity and matter~\cite{Doneva:2022ewd}.
Also, a nonminimal coupling to the Maxwell term  has induced   spontaneous scalarization of Reissner-Nordstr\"{o}m (RN) black hole in the Einstein-Maxwell-Scalar (EMS) theory~\cite{Herdeiro:2018wub}.
 A little difference is that  the former is hard to accommodate its extremal black hole with scalar hair, while the latter can possess  scalarized extremal  black hole when extending  dyonic (electric and magnetic charges) RN black holes~\cite{Astefanesei:2019pfq}.

Finding a scalarized extremal black hole with secondary scalar hair seems to be  a difficult task in the analytical and numerical ways.
It is known  that the BBMB black hole was found from the Einstein-conformally coupled scalar theory~\cite{Bocharova:1970skc,Bekenstein:1974sf}.
Even though  the scalar hair of $\phi(r)=\frac{m}{r-m}$ with black hole mass $m$  belongs to secondary scalar hair~\cite{Charmousis:2015aya}, it  blows up at the horizon.
Scalarization of RN black holes was investigated at the approach to extremality  by introducing the Einstein-Maxwell-Gauss-Bonnet-scalar theory with a quadratic scalar coupling to GB term~\cite{Brihaye:2019kvj}.
 It was shown that two branches of scalarized black holes appeared  when considering the  near-horizon geometry  of  extremal black holes.
It is worth noting that there were some examples for the  entropy function approach~\cite{Sen:2005wa} to find scalarized extremal black holes~\cite{Marrani:2017uli,Marrani:2022hva,Myung:2026ook}.
Here, we wish  to point out  its weakness: it is not easy  to find extremal black holes with secondary scalar hair because it used an attractor mechanism with constant scalar.

Recently, the EMS theory with two different scalar couplings to two  U(1) fields was used to find  two scalarized extremal black holes with constant scalar hair~\cite{Chew:2026clh}.
This is a similar case to the dyonic EMS theory which was employed firstly for finding scalrized extremal black holes~\cite{Astefanesei:2019pfq,Chen:2026olq}.

In the present work, we wish to investigate scalarized extremal black holes with scalar hair in the generic Einstein-Euler-Heisenberg-scalar (EEHS) with two scalar couplings $g_{\rm exp}(\phi)$ and $g_{\rm poly}(\phi)$ to the Maxwell term. When adopting spontaneous scalarization with $\phi_\infty=0$, we find infinite branches of scalarized non-extremal EEH black holes.
Here, it is found that there also exist infinite branches of scalarized extremal EEH black hole by introducing the Bertotti-Robinson geometry (BR: AdS$_2\times S^2$) as the near-horizon geometry of extremal EEH black holes. We will find a phenomenon of approaching all branches to the single critical branch at the approach to extremality.
Scalarization approach is employed to find scalarzied extremal black hole with constant secondary scalar hair for $\phi_\infty=\phi_h$ with horizon scalar $\phi_h$ when considering the polynomial coupling $g_{\rm poly}(\phi)$.
Furthermore,  this constant scalar solution is exactly recovered from the entropy function approach working on the BR geometry of extremal EEH black hole.
Importantly, we obtain scalarzied extremal EEH black hols with nonconstant scalar hair for the polynomial coupling $g_{\rm poly}$ when imposing the zero infinity scalar of $\phi_\infty=0$. Finally, the connection between scalarzied non-extremal black holes and extremal black holes with  constant and nonconstant scalar will be explored.

\section{Generic Einstein-Euler-Heisenberg-scalar theory } \label{sec:theory}

We start with the generic Einstein-Euler-Heisenberg-scalar theory ~\cite{Bakopoulos:2024hah}
\begin{eqnarray}
S=\frac{1}{16\pi}\int  d^4x\sqrt{-g}\Big[R-2\nabla^\mu\phi\nabla_\mu\phi-g(\phi)F^2-f(\phi)\left(2\gamma F^\mu_{~\nu} F^\nu_{~\rho} F^\rho_{~\delta} F^\delta_{~\mu}-\eta F^4\right)\Big],\label{action}
\end{eqnarray}
where $R$ denotes the scalar curvature, $g(\phi)$ and $f(\phi)$ are scalar coupling functions to the Maxwell term  ($F^2=F_{\mu\nu}F^{\mu\nu}$) and nonlinear electrodynamics (NED) terms with  $F^4=(F^2)^2$.
Here $\gamma$ and $\eta$ represent two scalar coupling constants to two  NED terms.
Varying the action \eqref{action} with respect to $g_{\mu\nu}$, $\phi$, and $A_\mu$, the three equations of motion are given by
\begin{eqnarray}
&&R_{\mu\nu}-\frac{1}{2}R g_{\mu\nu}=2{\partial _\mu}\phi{\partial_ \nu}\phi-g_{\mu\nu}{\partial }^\mu\phi {\partial }_\mu \phi+2T_{\mu\nu},\label{eqG}\\
&&\Box  \phi -\frac{1}{4}\frac{dg(\phi)}{d\phi}F^2-\frac{df(\phi)}{d\phi}\left(\frac{\gamma}{2} F^\mu_{~\nu} F^\nu_{~\zeta} F^\zeta_{~\delta} F^\delta_{~\mu}-\frac{\eta}{4} F^4\right)=0,\label{eqKG}\\
&&{\partial_ \mu}\Big[\sqrt{-g}\Big(g(\phi)F^{\mu\nu}+f(\phi)(4\gamma F^\mu_{~\kappa} F^\kappa_{~\lambda} F^{\nu\lambda}-2\eta F^2 F^{\mu\nu})\Big)\Big]=0,\label{eqMaxwell}
\end{eqnarray}
where $T_{\mu\nu}$ is the energy-momentum tensor  defined by
\begin{eqnarray}
T_{\mu\nu}&=&g(\phi)\Big(F^{\alpha}_\mu F_{\nu \alpha}-\frac{1}{4}g_{\mu\nu }F^2\Big)\nonumber\\
&&+f(\phi)\left ( 4\gamma F^\alpha _\mu F^\beta_\nu F^\eta_\alpha F_{\beta\eta} -\frac{\gamma}{2} g_{\mu\nu} F^\alpha_\beta F^\beta _\zeta F^\zeta _\delta F^\delta_\alpha -2\eta F^\xi_\mu F_{\nu\xi}F^2+\frac{\eta}{4}g_{\mu\nu} F^4\right).\nonumber
\end{eqnarray}

First of all, we briefly mention famous known black hole solutions obtained from the action \eqref{action}.
It has admitted the GMGHS black hole~\cite{Gibbons:1987ps,Garfinkle:1990qj} as an analytic solution when choosing $g(\phi)=e^{-2\phi}$ and $f(\phi)=0$ with  $\phi$ dilaton.
Considering $g(\phi)=e^{-2\phi}$ and $f(\phi)=-3\cosh(2\phi)-2$  led to the low-energy limit of  string theory and Lovelock-inspired gravity~\cite{Bakopoulos:2024hah}.
In this case, Eqs.\eqref{eqG}-\eqref{eqMaxwell} has shown a magnetically charged black hole solution with dilaton hair $\bar{\phi}(r)$ as
\begin{eqnarray}
 ds^2&=&-H(r) \,dt^2 + \frac{1}{H(r)} \, dr^2 + R^2(r) \,( d\theta^2 + \sin^2 \theta \, d\varphi^2),\\
H(r)&=&1-\frac{2 M}{r}+\frac{2\epsilon P^4}{r^3(r-P^2/M)^3}, \quad
R^2(r)=r\left(r-\frac{P^2}{M}\right),\nonumber\\
\bar{\phi} (r)&=&-\frac{1}{2}\ln \left(1-\frac{P^2}{M r}\right), \quad A_\varphi= P\cos\theta,
\end{eqnarray}
where $M$ and $P$ are the mass and magnetic charge with the Euler-Heisenberg (EH) paramter $\epsilon=\eta-\gamma$.
In the limit of $P\to 0$, one recovers Schwarzschild black hole.  
The polar perturbation analysis of this black hole was performed by computing dilaton and Zerilli quasinormal mode spectra, leading to a proof of its stability~\cite{Li:2026gqi}. 
For an exponential coupling of $g(\phi)=e^{\alpha\phi^2}$ and $f(\phi)=\{1,e^{\alpha \phi^2}\}$ with $\epsilon=0$~\cite{Myung:2020ctt}, spontaneous scalarization of nonextremal dyonic black holes gives a result similar to the EMS theory with the same coupling.

For the magnetically charged configuration with $g(\phi)=f(\phi)=1$, the hairless EEH black hole is given by~\cite{Bakopoulos:2024hah,Luo:2026srx}, while its dyonic generalization has recently been obtained in~\cite{Luo:2026ndd}.
In the present work, we restrict ourselves to the purely magnetic sector, for which the EEH black hole is given by
\begin{eqnarray}
&&ds^2_{\rm EEH}=\bar{g}_{\mu\nu}dx^\mu dx^\nu=-B(r) dt^2+\frac{dr^2}{B(r)} +r^2d\Omega^2_2,  \nonumber \\
&& B(r)=1-\frac{2M}{r}+\frac{P^2}{r^2}-\frac{2\epsilon P^4}{5r^6}, \label{dRN-bh}\\
&& \bar{A}=P\cos\theta d\varphi. \nonumber
\end{eqnarray}
For the magnetic configuration, all field-strength invariants are given by
\begin{equation}
\bar{F}^2=\frac{2P^2}{r^4},\qquad \bar{F}^\mu_{~\nu} \bar{F}^\nu_{~\rho} \tilde{F}^\rho_{~\delta} \bar{F}^\delta_{~\mu}=\frac{2P^4}{r^8},\qquad \bar{F}^4=\frac{4P^4}{r^8}. \label{invariants}
\end{equation}
They depend only on $g^{\theta\theta}g^{\varphi\varphi}$ and are therefore independent of $g_{tt}$. 
In particular, we have $2\gamma \bar{F}^{\mu}{}_{\nu}\bar{F}^{\nu}{}_{\rho}\bar{F}^{\rho}{}_{\delta}\bar{F}^{\delta}{}_{\mu}-\eta \bar{F}^{4}=4(\gamma-\eta)P^{4}/r^{8}=-4\epsilon P^{4}/r^{8}$,showing explicitly that the two NED couplings enter only through the EH parameter $\epsilon=\eta-\gamma$.
Scalarized EEH black holes with a single horizon were obtained through spontaneous scalarization for the scalar couplings $g(\phi)=e^{-\alpha \phi^2}$, $f(\phi)=1$~\cite{Zhang:2025msi} and $f(\phi)=e^{-\alpha \phi^2}$~\cite{Zhang:2026bqu} with $\epsilon=0.3$. 
In these cases, we note that there is no restriction on the magnetic charge $P$ as $P\in[0,\infty]$.
 
\begin{figure*}[t!]
   \centering
    \includegraphics[width=0.45\textwidth]{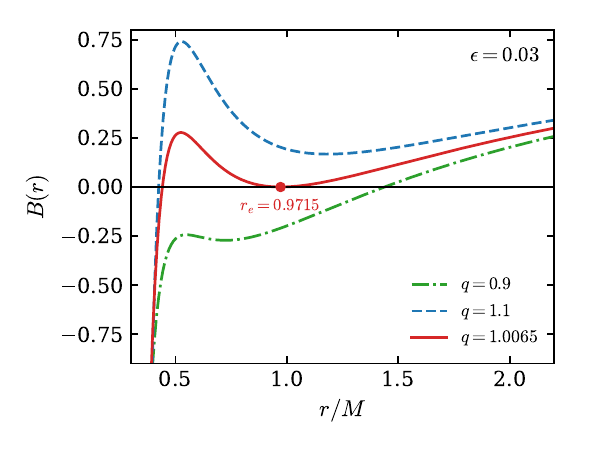}
   \includegraphics[width=0.45\textwidth]{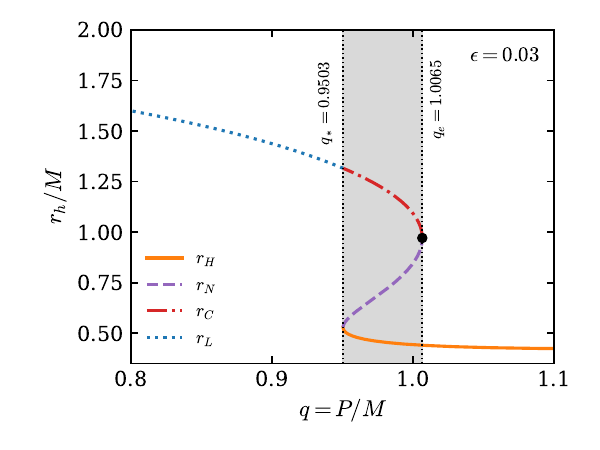}
\caption{(left) Metric function $B(r;M=1,q,\epsilon=0.03)$ at $q=0.9,\,1.0065,\,1.1$.
At $q_e=1.0065$, one real root and one double root remain, i.e.\ an extremal horizon.
(right) Four horizon branches (Low $r_{L}$, Cold $r_{C}$, Negative $r_{N}$, Hot $r_{H}$) as functions of $q$ at $\epsilon=0.03$.
The shaded region is the narrow window $q\in[q_*=0.95,q_e]$ in which three horizons coexist, its upper end is the extremal point.}
\label{fig:background}
\end{figure*}

Here we work instead in the regime $\epsilon\le0.08$, where Eq.~\eqref{dRN-bh} possesses multiple horizons. 
This choice is a necessary step to allow a degenerate horizon. 
Throughout the paper, we fix $\epsilon=0.03$ and $M=1$, and write $q=P/M$.

The left panel of Fig.~\ref{fig:background} shows how a degenerate (extremal) horizon forms as $q$ grows. 
For $\epsilon<0.08$, the horizon equation $B(r)=0$ has at most three positive roots, whose number is controlled by the two turning points of the horizon curve, $q_*=0.9503\simeq 0.95$ at $r_s\simeq0.529M$ and $q_e=1.00648\simeq 1.0065$ at $r_e\simeq0.9715M$. 
For $q<q_*$, there is a single horizon and  the outer branch reduces to the Schwarzschild radius $r=2M$ at $q=0$. 
At $q=q_*$, the inner pair of $r_N$ and  $r_H$ is born at a saddle node with double root $r\simeq0.529M$. 
For $q_*<q<q_e$, three horizons coexist, $r_H<r_N<r_C$. 
At $q=q_e$, the branches $r_N$ and $r_C$ merge into the degenerate horizon $r_e$, the genuine extremal point, because $r_N$ plays the role of the inner horizon of $r_C$. 
For $q>q_e$, only the inner branch $r_H$ remains. 
Following the thermodynamic classification of Ref.~\cite{Myung:2025zxu}, the branches are labeled $r_L$ (low temperature), $r_C$ (cold), $r_N$ (negative) and $r_H$ (hot). 
In this notation, $r_L$ and $r_C$ are not two independent horizons but two segments of the same outer branch, split at the turning point $q_*$, with their charge ranges $q_L\in[0,q_*]$ and $q_C\in[q_*,q_e]$. 
The ranges of the inner pair are $q_N\in[q_*,q_e]$ and $q_H\in[q_*,\infty)$. 
We note that $r_L$ and $r_H$ are connected to each other for $\epsilon>0.08$, but they are disconnected for $\epsilon<0.08$.

For later use, we record the data of the degenerate horizon. 
Imposing $B=B'=0$ and eliminating $M$ leaves
\begin{equation}
r_e^2=P^2-\frac{2\epsilon P^4}{r_e^4},\qquad
N_2\equiv\frac{1}{2}B''(r_e)=\frac{1}{r_e^2}\Big[1-\frac{4\epsilon P^4}{r_e^6}\Big], \label{loweq-3}
\end{equation}
so that $B(r)\simeq N_2 (r-r_e)^2$ near $r=r_e$. 
For $\epsilon=0.03$ and $M=1$, this gives $q_e=1.0064849$, $r_e=0.9715475$ and $N_2=0.9042983$ precisely.

\section{Scalarized C-horizon black hole}\label{sec:cold}
To carry out scalarization, we use the two-type  scalar couplings for $g(\phi)$ with $f(\phi)=1$ as 
\begin{equation}
g_{\rm exp}(\phi)=e^{-\alpha\phi^2},\qquad g_{\rm poly}(\phi)=1-\alpha\phi^2+\beta\phi^4, \label{couplings}
\end{equation}
which share $g(0)=1,~g'(0)=0,~g''(0)=-2\alpha$. 
The first two conditions guarantee that the hairless EEHBH (\ref{dRN-bh}) remains a solution of the full system, while the third yields a tachyonic mass for $\alpha>0$. 
The polynomial form of $g_{\rm poly}(\phi)$ plays the role  being familiar from the self-interaction studies of Refs.~\cite{Macedo:2019sem,Blazquez-Salcedo:2018jnn}, and we shall see in Sec.~\ref{sec:extremal} that it is exactly what makes a nontrivial extremal solution possible.

Before addressing extremality, we construct scalarized non-extremal black holes on the C horizon over the full window $q\in[q_*,q_e]$.  The reason why we choose C-horizon is that it contains an extremal black hole at $q=q_e$. 
The previous study of this system was restricted to the case of $q=1$~\cite{Guo:2026qib}. 
The $q$-dependence turns out to be essential, because the extremal point $q=q_e$ is where all scalarized branches accumulate.

We consider the metric and fields as~\cite{Herdeiro:2018wub}
\begin{eqnarray}\label{nansatz}
&&ds^2_{\rm SBH}=-N(r)e^{-2\delta(r)}dt^2+\frac{dr^2}{N(r)}+r^2(d\theta^2+\sin^2\theta d\varphi^2), \nonumber \\
&&N(r)=1-\frac{2m(r)}{r},\quad \phi=\phi(r),\quad A=P\cos\theta d\varphi
\end{eqnarray}
with $f(\phi)=1$. 
Plugging Eq.~\eqref{nansatz} into Eqs.~\eqref{eqG} and \eqref{eqKG}, one finds three equations
\begin{eqnarray}
&&r^2[N(r)+rN'(r)-1]+g(\phi)P^2-\frac{2\epsilon P^4}{r^4}
+r^4N(r)\phi'^2(r)=0, \label{neom1}\\
&&\delta'(r)=-r\phi'^2(r), \label{neom2}\\
&&[e^{-\delta(r)} r^2 N(r)\phi'(r)]'
-\frac{g'(\phi)e^{-\delta(r)}P^2}{2r^2}=0. \label{neom3}
\end{eqnarray}
Note that the length dimension of $\epsilon$ is given by 2, so that it is always measured in units of $M^2$.
As anticipated below Eq.~\eqref{invariants}, $\delta(r)$ is absent from Eq.~\eqref{neom1} and is factored out of Eq.~\eqref{neom3}.
Using Eq.~\eqref{neom2} to eliminate $\delta'(r)$, the scalar equation becomes
\begin{equation}
[r^2N(r)\phi'(r)]'+r^3N(r)\phi'^3(r)-\frac{g'(\phi)P^2}{2r^2}=0. \label{neom3p}
\end{equation}
Hence, Eqs.~\eqref{neom1} and \eqref{neom3p} form a closed system for $(N,\phi)$, while $\delta(r)$ will be determined by direct integration of Eq.~\eqref{neom2} with the boundary condition $\delta(\infty)=0$.

\subsection{Tachyonic instability and bifurcation points}\label{sec:linear}

\begin{figure*}[t!]
   \centering
   \includegraphics[width=0.45\textwidth]{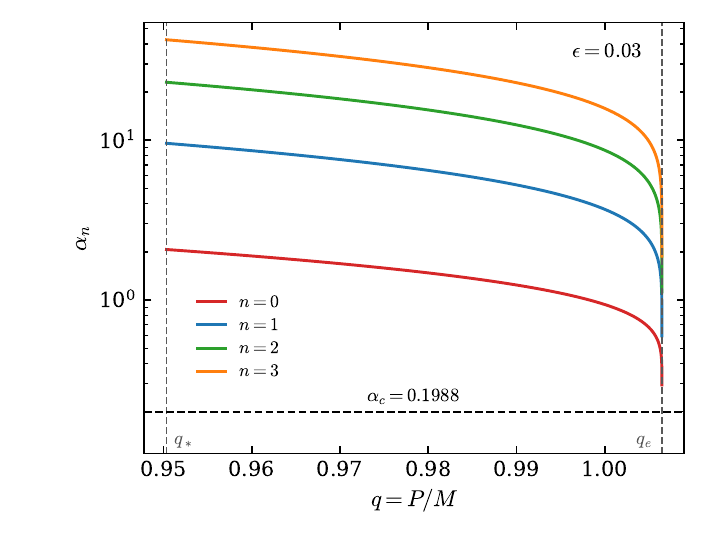}
   \includegraphics[width=0.45\textwidth]{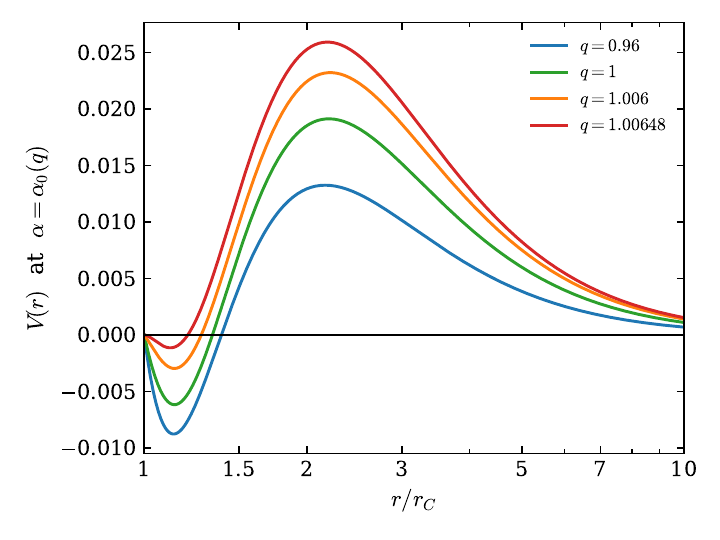}
\caption{(left)  Existence curves $\alpha_n(q)$ for  the scalarzied  C-horizon EEHBH with $\epsilon=0.03$ and  $n=0,1,2,3$. 
The horizontal dashed line denotes the critical branch of $\alpha_c=0.19884$ of Eq.(\ref{alpha-c}), while the two vertical dashed lines represent the ends of  $q_*$ and $q_e$ for  the triple-horizon window. 
All curves terminate on $\alpha_c$ at $q=q_e$.
(right) Effective potential $V(r)$ of Eq.(\ref{pot}) evaluated on the $n=0$ branch  of $\alpha=\alpha_0(q)$  for the same $\epsilon=0.03$ and $q=0.96,~1.00,~1.006,~1.00648$.}
\label{fig:linear}
\end{figure*}

Linearizing Eq.~\eqref{neom3p} around $\phi=0$ on the background \eqref{dRN-bh} and decomposing in spherical harmonics gives a radial scalar equation 
\begin{equation}
\big[r^2B(r)\phi'\big]'+\Big[\frac{\alpha P^2}{r^2}-\ell(\ell+1)\Big]\phi=0, \label{lin-eq}
\end{equation}
which is the same linearized equation for both couplings in Eq.~\eqref{couplings}. Equivalently, with $\phi=\psi/r$ and $dr_*=dr/B$, we can rewrite the scalar equation in terms of  the Schr\"odinger form
\begin{equation}
\frac{d^2\psi}{dr_*^2}=V(r)\psi,\qquad
V(r)=B\Big[\frac{\ell(\ell+1)}{r^2}+\frac{B'}{r}+\mu^2_{\rm eff}\Big],\qquad
\mu^2_{\rm eff}=-\frac{\alpha P^2}{r^4}. \label{pot}
\end{equation}
The negative region of $V$ generated by $\mu^2_{\rm eff}<0$ is displayed in the right panel of Fig~\ref{fig:linear}.
A bifurcation point corresponds to a value of coupling constant  $\alpha$ at which the hairless background admits a static  scalar zero mode that is regular on the horizon and decays at infinity.
At that value, the linearized operator has a nontrivial kernel and a branch of hairy solutions emerges  when the standard Lyapunov-Schmidt argument was  used for the scalarization~\cite{Doneva:2017bvd,Doneva:2022ewd,Herdeiro:2018wub}.
Eq.~\eqref{lin-eq} is already a  formally self-adjoint Sturm-Liouville form as 
\begin{equation}
\big(p\,\phi'\big)'+\big[\alpha\,w-\ell(\ell+1)\big]\phi=0,\qquad p=r^2B\ge0,\qquad w=\frac{P^2}{r^2}>0,
\end{equation}
so that the coupling constant $\alpha$ becomes itself the eigenvalue and the weight $w$ is strictly positive throughout the domain.
Treating a nonminimal coupling constant $\alpha$  as a Sturm-Liouville eigenvalue in this way is not new, and has been used both for the condensation threshold of holographic superconductors~\cite{Siopsis:2010uq} as well as  for the discrete spectrum of Gauss-Bonnet coupling in the  scalarization of Schwarzschild black hole~\cite{Hod:2019vut}.

Introducing a compact coordinate with $x=1-\frac{r_C}{r}\in[0,1]$, Eq.~\eqref{lin-eq} becomes the unit-weight form
\begin{equation}
-\big(B\,\phi_x\big)_x=\left(\lambda-\ell(\ell+1)\frac{r^2}{r_C^2}\right)\phi,\qquad \lambda=\frac{\alpha P^2}{r_C^2}. \label{SL}
\end{equation}
In this case, smoothness of the scalar at the infinity selects the constant solution.
The equivalent statement requires  the vanishing of the flux, $B\phi_x\to0$. 
At $x=1$, the condition for the mode to decay gives the Dirichlet condition $\phi(1)=0$, leading to  $\phi_\infty=0$. 
Under these conditions, the operator of  ``$-\partial_x(B\partial_x)$'' is self-adjoint, bounded below and has a compact resolvent, so that its spectrum $\lambda_0<\lambda_1<\cdots$ is real, simple, and discrete.
By employing the Sturm oscillation theorem, the $n$-th eigenfunction has exactly $n$ nodes in $(0,1)$~\cite{Zettl:2005}. 
Therefore,  the bifurcation points (existence curves) for infinite branches can be obtained analytically as 
\begin{equation}
\alpha_n(q)=\frac{r_C^2(q)}{P^2}\,\lambda_n(q),\qquad n=0,1,2,\dots .\label{alpha-n}
\end{equation}
From Eq.~\eqref{SL}, we know the $s(\ell=0)$-mode gives the lowest threshold scalar  and we restrict to the  $\ell=0$ from now on.

Solving Eq.~\eqref{SL} by a cell-centered finite-volume discretization gives the existence curves for $n=0,1,2,3$ branches shown in the left panel of Fig.~\ref{fig:linear}. 
They decrease monotonically with $q$, so that a colder black hole is easier to be scalarized. 
The fundamental ($n=0$) curve [$\alpha_{0}(q)=\alpha_{\rm th}(q)$] falls from $\alpha_0=2.0587$ at $q=0.951$ to $\alpha_0=0.3565$ at $q=1.00648$, and the entire spectrum of four branch points ($\alpha_0(q)<\alpha_1(q)<\alpha_2(q)<\alpha_3(q)$) approaches  to zero as $q\to q_e$.

\begin{table}[h!]
\centering
\caption{Semiclassical phase integrals $\int\!\sqrt{-V}\,dr_*$ evaluated on the threshold coupling $\alpha=\alpha_0(q)$ for $\epsilon=0.03$ and $M=1$.}
\label{tab:wkb}
\begin{tabular}{lcccc}
\hline
$q$ & $0.96$ & $1.00$ & $1.006$ & $1.00648$ \\
\hline
$\alpha_0(q)$ & $1.886870$ & $0.937609$ & $0.560122$ & $0.356479$ \\
$\min V$ & $-8.771\times10^{-3}$ & $-6.173\times10^{-3}$ & $-2.968\times10^{-3}$ & $-1.142\times10^{-3}$ \\
$\int\sqrt{-V}\,dr_*$ & $0.9541$ & $1.2878$ & $1.7744$ & $2.7098$ \\
\hline
\end{tabular}
\end{table}

The mechanism behind that decrease is visible in the effective potential $V(r)$, as shown in the right panel of Fig.~\ref{fig:linear}, which is evaluated on the threshold coupling $\alpha=\alpha_0(q)$. 
Outside the horizon, the bracket in Eq.~(\ref{pot}) is dominated by $\mu^2_{\rm eff}$, so we have $V(r)<0$ there. 
The zero mode is therefore bound in a negative potential well attached to the horizon. 
What changes with $q$ is the shape of this well. 
Along the threshold, the well becomes steadily shallower, its depth falling by almost an order of magnitude across the window, and yet the natural measure of the semiclassical phase across it ( $\int\!\sqrt{-V}\,dr_*$) grows as $q\to q_e$ (see Table~\ref{tab:wkb}). 
Since the phase integral grows while the well becomes shallower, the well must widen in the extremal point at $r=r_*$. 
Near the extremal point $q=q_e$, it extends over an increasingly larger interval and a shallow potential is sufficient to confine the mode because the mode gets more room. 
This describes  the geometric meaning of the decrease of $\alpha_n(q)$ as a function of $q$.   Sec~\ref{sec:accum} will turn this observation into deriving the quantitative law~Eq.~\eqref{accum}.

\subsection{Nonlinear scalarized C-horizon black holes}\label{sec:nonlinear}

To obtain the nonlinear solutions, we integrate Eqs.~\eqref{neom1} and~\eqref{neom3p} outward from the horizon $r_h$, defined by $N(r_h)=0$ by introducing a shooting method. 
Near the horizon in this theory, the fields admit the expansions with $\rho= r-r_h$ as 
\begin{eqnarray}
&&N(r)=N_1\rho+N_2\rho^2+N_3\rho^3+\cdots, \label{aps-0}\\
&&\delta(r)=\delta_0+\delta_1\rho+\cdots,\label{aps-1}\\
&&\phi(r)=\phi_h+\phi_1\rho+\cdots,\label{aps-2}
\end{eqnarray}
in which $\phi_h=\phi(r_h)$ is the horizon scalar serving as the continuation parameter and $\delta_0$ is fixed afterwards by asymptotic flatness. 
Inserting Eqs.~\eqref{aps-0}-\eqref{aps-2} into Eqs.~\eqref{neom1}-\eqref{neom3}, the lowest order gives the three horizon relations
\begin{equation}
N_1=\frac{1}{r_h^3}\Big[r_h^2-g(\phi_h)P^2+\frac{2\epsilon P^4}{r_h^4}\Big],\qquad
r_h^4\,N_1\,\phi_1=\frac{g'(\phi_h)P^2}{2},\qquad
\delta_1=-r_h\phi_1^2 . \label{hor-exp}
\end{equation}
These also hold at a degenerate horizon  and the middle relation distinguishes the two  coupling cases. 
Since the surface gravity is given by  $N_1e^{-\delta_0}/2$, so that the Hawking temperature is $T_H=N_1e^{-\delta_0}/4\pi$, and  a horizon is non-degenerate precisely when $N_1>0$. 
The middle relation then determines the horizon scalar gradient
\begin{equation}
\phi_1=\frac{g'(\phi_h)P^2}{2r_h^4N_1} ,
\end{equation}
which implies that $\phi_1$  is nonzero whenever $g'(\phi_h)\ne0$.

On the other hand, the near-horizon solution should be  matched with  the asymptotic form
\begin{eqnarray}
&&N(r)=1-\frac{2M}{r}+\frac{g(\phi_\infty)P^2+Q_s^2}{r^2}+\cdots, \\
&&\delta(r)=\frac{Q_s^2}{2r^2}+\cdots,\\
&&\phi(r)=\phi_\infty+\frac{Q_s}{r}+\cdots, \label{asyp-3}
\end{eqnarray}
where $Q_s$ and $\phi_\infty$ denote the scalar charge and asymptotic scalar, in addition to the ADM mass $M$ and magnetic charge $P$. 
Here, it is important to note that the physical branch is fixed by $\phi_\infty=0$. 
Setting $M=1$ and $P=q$, implying  that  the charge-to-mass ratio is  equal to the prescribed $q$. 
This can be  achieved by the scaling freedom: $r\to\tilde{\lambda} r$, $M\to\tilde{\lambda} M$, $P\to\tilde{\lambda} P$, $\epsilon\to\tilde{\lambda}^2\epsilon$  under which the scalar $\phi$ is invariant. 
Now, the physical dimensionless parameters are given by  $(q,\ \epsilon/M^2,\ \alpha,\ \beta)$.

\begin{figure*}[t!]
   \centering
   \begin{minipage}[t]{0.32\textwidth}
       \centering
       \includegraphics[width=\linewidth]{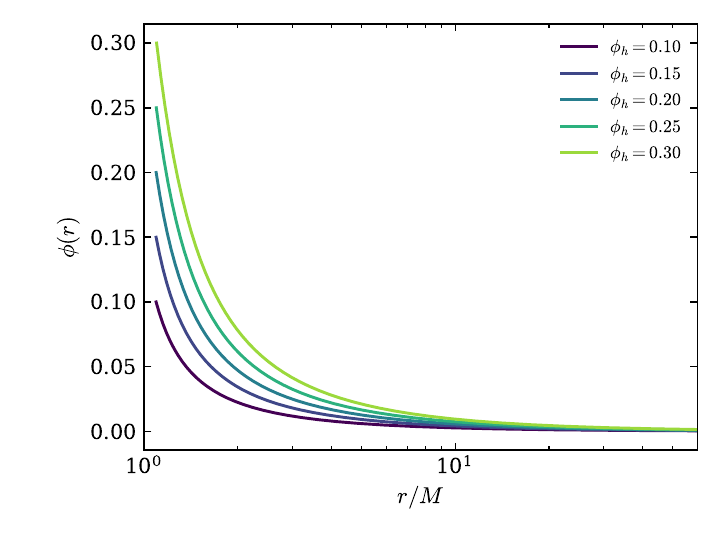}
   \end{minipage}
   \begin{minipage}[t]{0.32\textwidth}
       \centering
       \includegraphics[width=\linewidth]{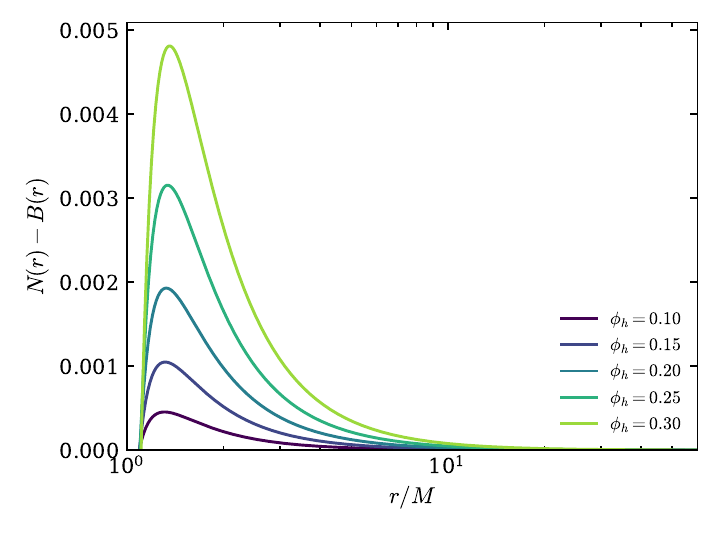}
   \end{minipage}
   \begin{minipage}[t]{0.32\textwidth}
       \centering
       \includegraphics[width=\linewidth]{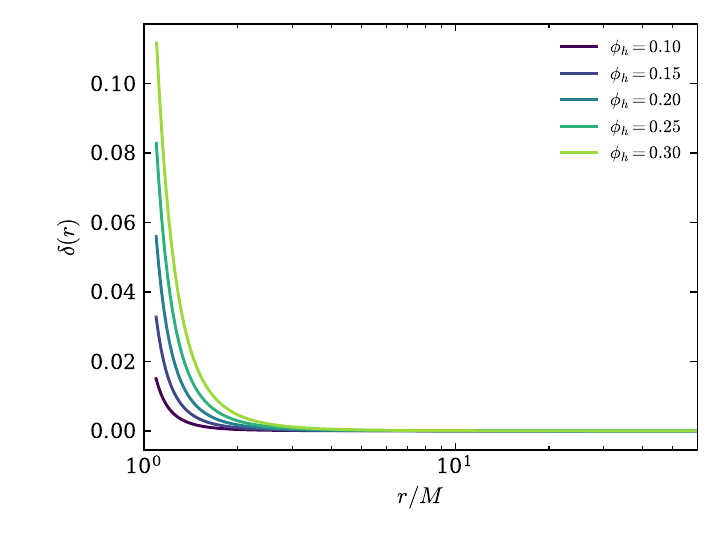}
   \end{minipage}
\caption{Profiles for nonlinear scalarized C-horizon black holes with  $g_{\rm exp}=e^{-\alpha\phi^2}$: (left) scalar hair  $\phi(r)$,  (middle) metric deviation $N(r)-B(r)$ from the hairless EEHBH with the same $(M,P)$, and (right)  $\delta(r)$ for $q=1$ and  $\epsilon=0.03$, taken at five points of horizon scalar  $\phi_h$ along the $n=0$ branch.}
\label{fig:profiles}
\end{figure*}

\begin{figure*}[t!]
   \centering
    \includegraphics[width=0.45\textwidth]{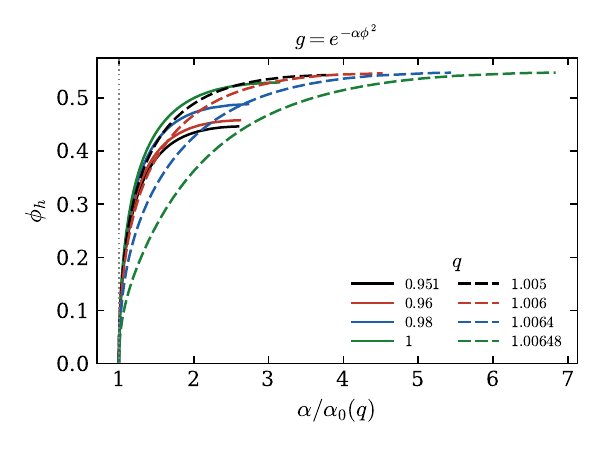}
    \includegraphics[width=0.45\textwidth]{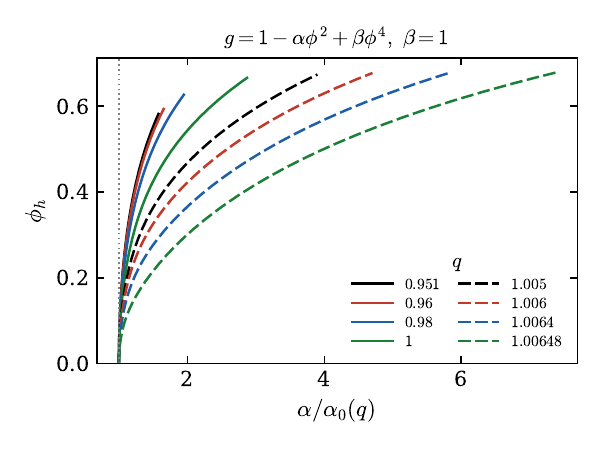}
\caption{Nonlinear scalarized C-horizon black holes with $\epsilon=0.03$. 
$\phi_h$ versus $\alpha/\alpha_0(q)$ for $g_{\rm exp}=e^{-\alpha\phi^2}$ (left) and $g_{\rm poly}=1-\alpha\phi^2+\beta\phi^4$ with $\beta=1$ (right). The exponential branches end at a turning point where $d\phi_h/d\alpha$ approaches zero, while the polynomial ones end where $g_{\rm poly}(\phi_h)\to0$.}
\label{fig:branches}
\end{figure*}

Representative profiles at $q=1$ are displayed in Fig.~\ref{fig:profiles} for $g(\phi)=e^{-\alpha \phi^2}$. 
The scalar is monotonic and node-free on the $n=0$ branch, while the metric backreactions of  $N(r)-B(r)$ and $\delta(r)$  are confined to the near-horizon region. 
Fig.~\ref{fig:branches} collects the resulting branches for both couplings in Eq.~\eqref{couplings} with $\beta=1$. 
Each branch emerges from $\phi_h=0$ exactly at $\alpha=\alpha_0(q)$ obtained independently from Eq.~\eqref{alpha-n}, and $\alpha$ increases with $\phi_h$, so that scalarized C-horizon black holes exist for $\alpha>\alpha_0(q)$. 
The two couplings are indistinguishable near the bifurcation point but they separate at larger $\phi_h$. 

At this stage, we note that the two couplings also terminate for different reasons. 
For $g=e^{-\alpha\phi^2}$, the branch reaches a maximal horizon value at which  $d\phi_h/d\alpha$ approaches zero, so that $\phi_h$ ceases to be a good parameter there and the branch turns over. 
For $g_{\rm poly}=1-\alpha\phi^2+\beta\phi^4$, the branch instead ends when $g(\phi_h)$ approaches zero, where the Maxwell kinetic term would change sign. 
Since $\phi$ takes a maximum value on the horizon, $g(\phi_h)>0$ is required. 
In both cases, $\phi_h^{\rm max}$ grows with $q$ over most of the window, but the colder black holes support the larger horizon scalar.

\begin{figure*}[t!]
   \centering
    \includegraphics[width=0.45\textwidth]{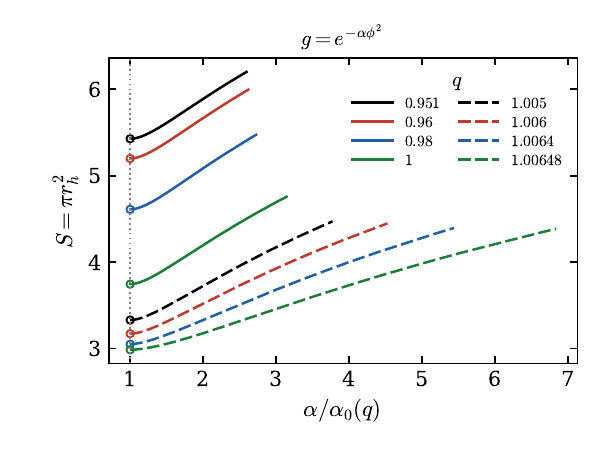}
    \includegraphics[width=0.45\textwidth]{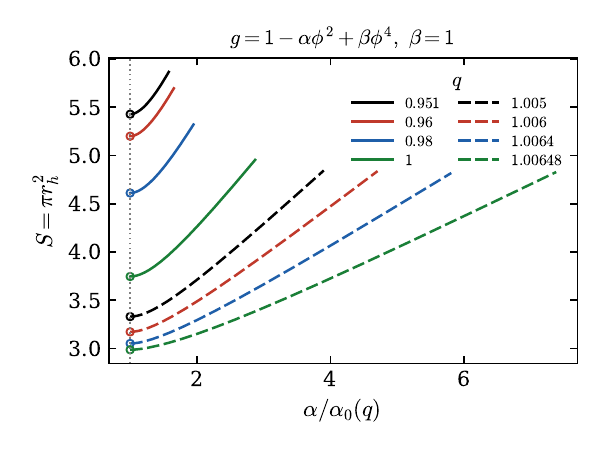}\\[2ex]
    \includegraphics[width=0.45\textwidth]{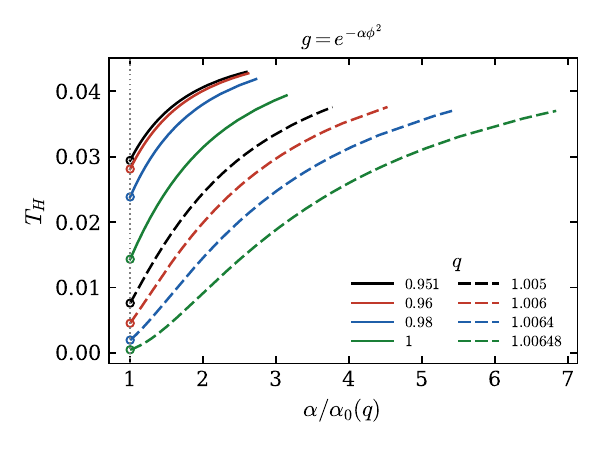}
    \includegraphics[width=0.45\textwidth]{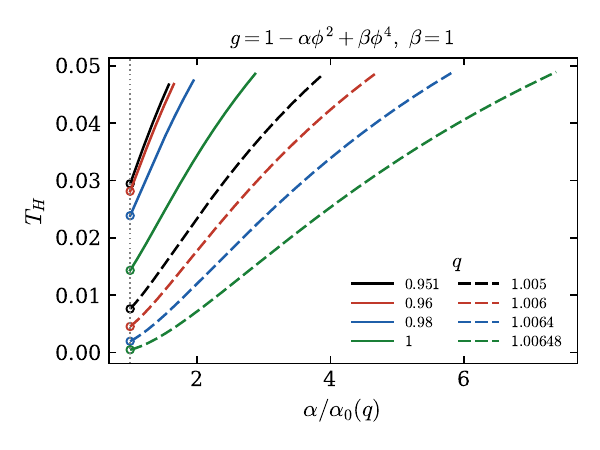}
\caption{Entropy $S=\pi r_h^2$ (top) and Hawking temperature $T_H$ (bottom) of the scalarized C-horizon black holes with $\epsilon=0.03$ and $M=1$.   (left)  $g_{\rm exp}=e^{-\alpha\phi^2}$ and  (right) $g_{\rm poly}=1-\alpha\phi^2+\beta\phi^4$ with $\beta=1$. The open circles on each curve represent those of the hairless EEHBHs with the same $(M,q)$ [at bifurcation points of $\alpha=\alpha_0(q)$ where the branch starts].}
\label{fig:thermo}
\end{figure*}

At fixed $(M,q)$, the scalarized solutions carry a larger horizon area than the hairless EEHBH: $S>S_{\rm EEH}=\pi r_C^2$, as shown in the upper panels of Fig.~\ref{fig:thermo}.  So, they are entropically preferred wherever they exist, in accordance with the EMS case~\cite{Herdeiro:2018wub}. 
We have verified branch by branch, for all selected charges and both couplings, that the excess is positive and grows monotonically from zero at the bifurcation point to its terminal value.

The temperature behaves in the same way, in the lower panels of Fig.~\ref{fig:thermo}:  $T_H>T_{\rm EEH}$ everywhere and monotonically increasing. 
The effect is most pronounced at $q\to q_e$, where the hairless reference temperature vanishes while the scalarized branch immediately acquires a finite temperature $T_H$, so that the ratio $T_H/T_{\rm EEH}$ grows without bound. 
Importantly, we observe that the scalar hair therefore lifts the near-extremal black hole off zero temperature. At a given $\alpha/\alpha_0(q)$, both $S$ and $T_H$ decrease with $q$, whereas the branch at $q=q_e$ is regarded as the coldest temperature and the smallest entropy of the eight selected charges.

\subsection{Approaching extremality: $q\to q_e$}\label{sec:accum}

It is important to  know  what happens to scalarization as the C-horizon merges with the N-horizon. Even though this was discussed in the Einstein-Gauss-Bonnet-Maxwell-scalar theory~\cite{Brihaye:2019kvj}, the process is not complete. 
In the language of the horizon expansion (\ref{aps-0}), this corresponds to  the limit $N_1\to0$ at fixed $N_2$. 
The linear term of $N$ ($N_1$)  disappears and it starts quadratically, as is shown in  $B\simeq N_2(r-r_e)^2$ with $N_2=\tfrac12B''(r_e)$ given by Eq.~\eqref{loweq-3}. 
The near-horizon geometry is then described by  the Bertotti-Robinson spacetime AdS$_2\times S^2$~\cite{Bertotti:1959pf,Robinson:1959ev},
\begin{equation}
ds^2\simeq -N_2\rho^2 dt^2+\frac{d\rho^2}{N_2\rho^2}+r_e^2d\Omega_2^2,\qquad \rho=r-r_e.
\end{equation}
The AdS$_2$ mass of the $\ell$-th multipole takes the form $m^2_{\rm AdS_2}=\mu^2_{\rm eff}+\ell(\ell+1)/v_1$, and the Breitenlohner-Freedman (BF) bound of  $m^2_{\rm AdS_2}v_0\ge-1/4$~\cite{Breitenlohner:1982jf} is violated for
\begin{equation}
\alpha>\alpha_c(\ell,P)=\frac{r_e^4}{P^2}\Big[\frac{\ell(\ell+1)}{r_e^2}+\frac{N_2}{4}\Big], \label{alpha-c}
\end{equation}
where $\alpha_c(\ell,P)$ denotes the critical coupling constant (critical branch).
For $\epsilon=0$ case of the extreme RN black hole, one has $r_e=P=M$ and $N_2=1/r_e^2$, giving exactly $\alpha_c(\ell=0)=0.25$. 
For our case of $\epsilon=0.03$, Eq.~\eqref{alpha-c} gives $\alpha_c=0.19884$ at $\ell=0$ and $2.06240$ at $\ell=1$, confirming that the $s(\ell=0)$-wave sets the threshold. 
Here, one finds that the Euler-Heisenberg terms therefore lower the extremal scalarization threshold below its RN value, which means that extremal EEHBH is easier to be  scalarized than extremal RN black hole. However, this describes  the AdS-tachyonic instability. 
\begin{figure*}[t!]
   \centering
    \includegraphics[width=0.95\textwidth]{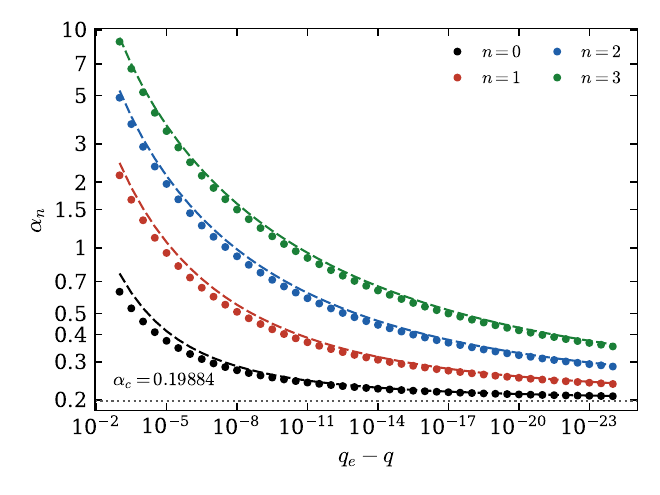}
\caption{Modified existence curves  $\alpha_n(q_e-q)$  ranging from $q_e-q=10^{-3}$ down to $10^{-24}$  with four branches $n=0,1,2,3$. Dots denotes  the numerical eigenvalues of Eq.\eqref{alpha-n}, dashed curves show the accumulation law Eq.\eqref{accum}, and the dotted line represents  the critical branch  $\alpha_c$ from Eq.(\ref{alpha-c}), onto which the whole spectra approach.}\label{fig:accum}
\end{figure*}

The approach of $\alpha_n(q)$ to $\alpha_c$ can also be obtained analytically.
Near extremality, we have  $B\simeq N_2(r-r_C)(r-r_N)$ and define
\begin{equation}
\sigma=\frac{r_C-r_N}{2},\qquad r_0=\frac{r_C+r_N}{2},\qquad x=\frac{r-r_0}{\sigma}\ (\ge1).
\end{equation}
One has $B=N_2\sigma^2(x^2-1)$ and $d/dr_*=N_2\sigma(x^2-1)\,d/dx$, so that inside the infinite throat, where $V\simeq B\,\mu^2_{\rm eff}$, the static scalar equation~\eqref{pot} reduces to the Legendre equation
\begin{equation}
\big[(x^2-1)\psi'\big]'+\Lambda\psi=0,\qquad \Lambda=-\mu^2_{\rm eff}v_0=\frac{\alpha}{4\alpha_c},
\end{equation}
whose solutions oscillate with the AdS$_2$ exponent
\begin{equation}
\nu^2=\Lambda-\frac14=\frac{\alpha-\alpha_c}{4\alpha_c}. \label{nu}
\end{equation}
Real $\nu$ imposes  precisely the condition $\alpha>\alpha_c$, so that  the BF condition and oscillation criterion assess the same statement. 
With $x=\cosh s$, the AdS$_2$-throat reaches $s_{\rm max}={\rm arccosh}\,(x_{\rm max})\simeq\ln(2/\sigma)$, up to an additive constant of order unity which does not affect the leading behavior. 
The $n$-th  scalar zero mode therefore requires $\nu(\alpha_n)\ln(2/\sigma)\simeq(n+1)\pi$ and hence, leads to the modified existence curves for infinite branches
\begin{equation}
\alpha_n-\alpha_c\simeq \frac{4\pi^2(n+1)^2\alpha_c}{\ln^2(2/\sigma)}. \label{accum}
\end{equation}
which describes the branch points for the $s$-mode scalar around the BR geometry.
Since $\sigma\propto\sqrt{q_e-q}$ is close to the merger, this yields  ``$\frac{1}{\ln^2(q_e-q)}$"-law. 
It is worth noting that although the convergence is extremely slow, but the collapsing of all branch points $\alpha_n$ to the critical  value $\alpha_c$ is universal when approaching  its extremal point at $q=q_e$.
This suggests that at the extremality, one may find the single critical branch.

Because the law of Eq.\eqref{accum} is logarithmic, testing it requires reaching very small $q_e-q$.  For this purpose, we have to decouple the background computation from the eigenvalue problem. 
The background is computed in extended precision, with the degenerate-horizon data obtained when eliminating $q$ between $B=0$ and $B'=0$. 
As Fig.~\ref{fig:accum} shows, Eq.~\eqref{accum} is confirmed by the numerical eigenvalues over the whole range, where the dashed curves represent the law itself. 
Its  exponents of $\nu_n$ reconstructed through Eq.~\eqref{nu} are equally spaced with the predicted spacing $\pi/\ln(2/\sigma)$. 
Equivalently, we confirm that  the combination $\nu_n\ln(2/\sigma)/\pi$ tends to integers, like as $0.973,~1.947,~2.925,~3.906$ for $n=0,1,2,3$ at $q_e-q=10^{-24}$. 
We stress that both are sharp tests, since an incorrect $\alpha_c$ would immediately destroy the equal spacing. 

Finally, the $1/\ln^2(2/\sigma)$ approach means that, although the extremal threshold is universal, it can never be reached at any finite distance from the observer located outside the extremal black hole. 

\section{Constant scalar hair and the entropy function approach}\label{sec:extremal}

In the previous section, for $\phi_\infty=0$, we notified the presence of degenerate horizon of the hairless EEHBH as well as all branches of scalarized EEHBHs  accumulated to the critical branch at the approach to extremality.
We now construct scalarized extremal black holes for the polynomial coupling $g_{\rm poly}(\phi)$ with  $\phi_\infty=\phi_h$, starting from the same  Eqs.~\eqref{neom1}-\eqref{neom3}.

\subsection{Constant scalar hair from the analytic near-horizon expansion}\label{sec:constant}

A scalarized extremal black hole is a solution whose horizon expansion \eqref{aps-0}-\eqref{aps-2} has $N_1=0$, hence we have $T_H=0$, so that $N=N_2\rho^2+N_3\rho^3+\cdots$ with $\rho=r-r_+$, and $r_+$ the value of $r_h$ at degeneracy. 
This is not a new boundary condition, which mean that the three relations \eqref{hor-exp} are common to both cases, and everything below follows by reading them at $N_1=0$.

The first relation then becomes a condition on the horizon radius alone,
\begin{equation}
r_+^2=g(\phi_h)P^2-\frac{2\epsilon P^4}{r_+^4}, \label{loweq-1}
\end{equation}
which is Eq.~\eqref{loweq-3} with $P^2\to g(\phi_h)P^2$. 
For $g(\phi_h)=1$, it yields 
\begin{equation}
r_e(P,\epsilon)=\frac{1}{\sqrt{3}}\sqrt{P^2+\frac{P^4}{\xi^{1/3}}+\xi^{1/3}}\label{eEEH}
\end{equation}
with
\begin{equation}
\xi=P^6-27P^4\epsilon+3\sqrt{3}\sqrt{27P^8\epsilon^2-2P^{10}\epsilon}.
\end{equation}
Eq.(\ref{eEEH}) recovers the extremal horizon for the EEHBH, $r_e(P=1.0065,\epsilon=0.03)=0.9715=r_C(M=1,P=1.0065,\epsilon=0.03)=r_N(M=1,P=1.0065,\epsilon=0.03)$ as shown in right panel of Fig.~\ref{fig:background}. 
In this case, its entropy is given by
\begin{equation}
    \mathcal{E}_{\rm e}=\pi r_e^2(P,\epsilon). \label{entropyE}
\end{equation}
The third relation still reads $\delta_1=-r_+\phi_1^2$.

At a degenerate horizon, the left-hand side of the middle relation in Eq.~\eqref{hor-exp} vanishes identically. 
It no longer fixes $\phi_1$ but instead constrains the scalar itself,
\begin{equation}
g'(\phi_h)=0. \label{loweq-2}
\end{equation}
Expanding Eq.~\eqref{neom3p} one order further gives
\begin{equation}
\Big(2N_2r_+^2-\frac{g''(\phi_h)P^2}{2r_+^2}\Big)\phi_1=-\frac{g'(\phi_h)P^2}{r_+^3} ,\label{loweq-4}
\end{equation}
so that, once Eq.~\eqref{loweq-2} is imposed in Eq.~\eqref{loweq-4}, we have $\phi_1=\delta_1=0$ unless
\begin{equation}
N_2=\frac{g''(\phi_h)P^2}{4r_+^4} .
\label{phi1zero}
\end{equation}
Repeating the argument order by order, all higher coefficients vanish as well. 
Away from those exceptional values, $\phi\equiv\phi_h$ and $\delta\equiv\delta_0$ are constants. 
The exceptional values are not an artefact, they correspond to the resonances analyzed in Sec.~\ref{sec:frobenius}, and $n=1$ branch is exactly the case excluded in Eq.~\eqref{phi1zero}.

For $g'(\phi_h)=0$, Eqs.~\eqref{loweq-3} and \eqref{loweq-1} give the complete solution
\begin{eqnarray}
&&r_+(P,\epsilon,\phi_h)=\frac{1}{\sqrt{3}}\sqrt{g(\phi_h)P^2+\frac{g^2(\phi_h)P^4}{\zeta^{1/3}}+\zeta^{1/3}}, \label{Nff1} \\
&& \phi_1=\delta_1=0, \quad  N_2=\frac{1}{r_+^2}\Big[1-\frac{4\epsilon P^4}{r_+^6}\Big], \nonumber
\end{eqnarray}
where
\begin{equation}
\zeta=g^3(\phi_h)P^6-27P^4\epsilon+3\sqrt{3}\sqrt{27P^8\epsilon^2-2g^3(\phi_h)P^{10}\epsilon}.
\end{equation}
Asymptotic flatness of the constant-scalar configuration then fixes $\delta_0=0$, while $\phi\equiv\phi_h$ trivially gives $\phi_\infty=\phi_h$ and $Q_s=0$.

Condition~\eqref{loweq-2} is a genuine selection rule on the coupling function.
For $g=e^{-\alpha\phi^2}$, $g'=-2\alpha\phi\,e^{-\alpha\phi^2}$ vanishes only at $\phi_h=0$. 
This means that the exponential coupling admits no such nontrivial extremal solution.
In contrast, $g=1-\alpha\phi^2+\beta\phi^4$ has a nontrivial root
\begin{equation}
\phi_0=\sqrt{\frac{\alpha}{2\beta}},\qquad g_0\equiv g(\phi_0)=1-\frac{\alpha^2}{4\beta},
\end{equation}
requiring $\alpha<2\sqrt\beta$ to ensure $g_0>0$ and preserve the sign of the Maxwell term. 
This shows clearly  why the two couplings, indistinguishable at the level of the existence curves  $\alpha_n(q)$ in Sec.~\ref{sec:linear}, behave completely differently at extremality.

In the extremal  case, the spacetime is asymptotically flat, static and regular outside the horizon, with $T_H=0$ and $S=\pi r_+^2$.
It corresponds to  in fact the hairless extremal EEH black hole of the same theory with the magnetic charge and the NED parameter rescaled as  $P'\equiv\sqrt{g_0}P$ and $\epsilon'\equiv\epsilon/g_0^2$.
Because the constant scalar carries neither  independent charge nor geometric information,  its hair is secondary in a stronger sense than usual.

Thanks to $\phi_\infty=\phi_h\ne0$, this solution does not belong to the asymptotic vacuum with $\phi_\infty=0$ containing the hairless background \eqref{dRN-bh} and all branches of scalarized non-extremal EEH black holes constructed in Sec.~\ref{sec:cold}. 
Nevertheless, $\phi_h$ has a clear physical meaning. 
When writing the scalar equation as $\Box\phi=\partial_\phi U$ with $U(\phi,r)=g(\phi)P^2/(2r^4)$, one has $g''(0)=-2\alpha<0$ and $g''(\phi_0)=4\alpha>0$. 
Thus, $\phi=0$ is the local maximum of $U(\phi,r)$ which is  responsible for the tachyonic- and AdS-instability in Sec.~\ref{sec:linear}, while its minimum is located  at $\phi=\phi_0$. 
The constant scalar hair solution is thus regarded as the endpoint of the tachyonic instability, and it never  arises from  the bifurcation off the hairless solution.

\subsection{Entropy function approach}

The same solution can be recovered independently from the entropy function approach~\cite{Sen:2005wa,Astefanesei:2007bf,Sen:2007qy}, which works directly on the near-horizon geometry and thereby provides a check point without any of the preceding horizon expansion.

For this purpose, we choose the BR line element and the matter ansatz
\begin{equation}
  ds^2_{\rm BR}=v_0\left(-r^2dt^2+\frac{dr^2}{r^2}\right)+v_1(d\theta^2+\sin^2\theta d\varphi^2),
  \qquad \phi=\phi_h,\quad A=P\cos \theta d\varphi.
\end{equation}
This is not an independent ansatz, since $(v_0,v_1)$ are the near-horizon data $(1/N_2,r_+^2)$ of the degenerate horizon.
For this background, its non-vanishing components of curvature tensor are given by
\begin{eqnarray}
&&R_{\alpha\beta\gamma\delta}=-\frac{1}{v_0}\Big(g_{\alpha\gamma}g_{\beta\delta}-g_{\alpha\delta}g_{\beta\gamma}\Big), \quad {\rm for} \quad \alpha,\beta,\gamma, \delta=t,r, \\
&&R_{mnpq}=\frac{1}{v_1}\Big(g_{mp}g_{nq}-g_{mq}g_{np}\Big), \quad {\rm for} \quad m,n,p,q=\theta,\varphi, 
\end{eqnarray}
giving $R=-2/v_0+2/v_1$, while $F^2=2P^2/v_1^2$ and the quartic combination follows from Eq.\eqref{invariants} with $r^2\to v_1$.
Three parameters $\{v_0,v_1,\phi_h\}$ satisfy a set of algebraic relations obtained from Eqs.\eqref{eqG}-\eqref{eqMaxwell}. 
Integrating the Lagrangian density over $S^2$  with $\int d\theta d\varphi\sqrt{-g}=4\pi v_0v_1$  yields
\begin{eqnarray}
  {\cal L} &=& \frac{1}{16 \pi}\int d\theta d\varphi\sqrt{-g}
	\Big[ R-2\partial_\mu \phi \partial^\mu \phi-g(\phi_h) F^2+\frac{4\epsilon P^4}{v_1^4} \Big] \nonumber  \\
&=&\frac{1}{2}
\left[
 v_0-v_1-\frac{ v_0g(\phi_h)P^2}{v_1}+\frac{2v_0\epsilon P^4}{v_1^3}\right]. \label{act-E}
\end{eqnarray}
Now, we define the entropy function $\mathcal{E}$ by taking the above density as
\begin{eqnarray}
\mathcal{E}=
  -2\pi {\cal L}.
\end{eqnarray}
We note that no Legendre transformation is needed, because the magnetic charge is not defined through $\partial{\cal L}/\partial F_{rt}$.
Extremizing with respect to $\{v_0,v_1,\phi_h\}$ gives
\begin{eqnarray}
  \label{T1}
  \frac{\partial {\cal E}}{\partial v_0} & = & 0\,\,\,\rightarrow \,\,\,
	v_1= g(\phi_h)P^2-\frac{2\epsilon P^4}{v_1^2},
	\\
  \label{T2}
  \frac{\partial {\cal E}}{\partial v_1} & = & 0\,\,\,\rightarrow \,\,\,
	\frac{v_0}{v_1^2}\Big[g(\phi_h)P^2-\frac{6\epsilon P^4}{v_1^2}\Big]=1,
	\\
  \label{T3}
  \frac{\partial {\cal E}}{\partial \phi_h} & = & 0\,\,\,\rightarrow \,\,\,
	g'(\phi_h) = 0,
\end{eqnarray}
Eq.~\eqref{T1} is the cubic of Eq.~\eqref{loweq-1} and is solved by
\begin{eqnarray}
v_1(P,\epsilon,\phi_h)=\frac{1}{3}\Big[g(\phi_h)P^2+\frac{g^2(\phi_h)P^4}{\kappa^{1/3}}+\kappa^{1/3}\Big], \label{t1}
\end{eqnarray}
where
\begin{equation}
\kappa=g^3(\phi_h)P^6-27P^4 \epsilon +3\sqrt{3}\sqrt{27P^8\epsilon^2-2g^3(\phi_h)P^{10}\epsilon}.
\end{equation}
Combining Eq.~\eqref{T1} with Eq.~\eqref{T2} implies
\begin{equation}
\frac{1}{v_0}=\frac{1}{v_1}\Big[1-\frac{4\epsilon P^4}{v_1^3}\Big]. \label{t2}
\end{equation}
Comparing with the near-horizon results, Eqs.~\eqref{T3}, \eqref{t1} and \eqref{t2} are respectively \eqref{loweq-2}, \eqref{Nff1} and the expression for $N_2$ in \eqref{Nff1}, under the identification of $v_1=r_+^2$ and $N_2=1/v_0$ together with $\delta_0=0$.
The two derivations therefore agree term by term. 

The entropy is the extremal value of $\mathcal E$.
Using Eq.~\eqref{T1} to write $g(\phi_h)P^2=v_1+2\epsilon P^4/v_1^2$ and then Eq.~\eqref{t2} to write $4v_0\epsilon P^4/v_1^3=v_0-v_1$, the four terms of Eq.~\eqref{act-E} collapse to yield
\begin{equation}
\mathcal{E}_{\rm c\phi}(P,\alpha,\beta,\epsilon)=-\pi\Big[v_0-2v_1-\frac{4v_0\epsilon P^4}{v_1^3}\Big]=\pi v_1=\pi r_+^2.
\label{entropyC}
\end{equation}
This is precisely the Bekenstein-Hawking entropy $A/4$ for the horizon area $A=4\pi v_1$.
For the polynomial coupling, the constant hair takes the form 
\begin{equation}
    \phi_h=\phi_0=\sqrt{\frac{\alpha}{2\beta}} \label{h-scalar}
\end{equation}
with $g_0=1-\alpha^2/4\beta$, both determined entirely by the action parameters. Here we observe that the entropy of $\pi v_1(P,\epsilon,\phi_h)$ is less than  the hairless extremal entropy $\pi r_e^2(P,\epsilon)$ with $\alpha<2\sqrt{\beta}(g_0>0)$ at the same magnetic charge.
This difference may indicate the observable consequence of the constant secondary hair.

\section{Nonconstant scalar hair in the extreme EEHS theory}

Before we proceed, it is noted that the analytic expansion of Sec.~\ref{sec:constant} captures only a part of the local solution space at a degenerate horizon.
Such a horizon is a regularly singular point of the scalar equation and the full local analysis is necessary to reveal a normal scalar mode missed by the analytic ansatz.

\subsection{Frobenius exponents and resonance at the degenerate horizon}\label{sec:frobenius}

The expansions of Eqs.\eqref{aps-0}-\eqref{aps-2} constitute an analytic (integer-power) ansatz. 
Since the degenerate horizon is a regularly singular point of the scalar equation, it is natural to ask for finding  a complete local solution space. 
For this purpose, linearizing scalar equation~\eqref{neom3p} around the constant scalar hair  $\phi_h$ on the throat $B\simeq N_2\rho^2$ together with  the operator $\Box= N_2\rho^2\partial_\rho^2+2N_2\rho\partial_\rho$, one finds the indicial equation
\begin{equation}
\frac{\lambda(\lambda+1)}{v_0}=m^2,\qquad
m^2=\frac{g''(\phi_h)P^2}{2r_+^4},
\label{indicial}
\end{equation}
which indicates  the AdS$_2$ scaling relation for a massive scalar with mass $m$. 
Here and below, we take $s(\ell=0)$-wave, which is the mode that drives scalarization. 
For $\ell>0$, the mass term acquires the shift of  $\ell(\ell+1)/v_1$ additionally. 
Its roots are
\begin{equation}
\lambda_\pm=-\frac12\pm\sqrt{\frac14+m^2v_0},
\label{lambdapm}
\end{equation}
so that near the degenerate horizon, the scalar field behaves as
\begin{equation}
\phi=\phi_h+c\,\rho^{\lambda_\pm}+\cdots .
\label{frob}
\end{equation}
We note that the negative  branch of  $\lambda_-<-1$ diverges and is discarded, whereas the positive branch of $\lambda_+>0$ vanishes at the horizon ($\rho=0$) and is completely regular.
So, we choose $\lambda_+$ branch only. 
Its amplitude $c$ denotes  a free parameter which is hidden  to the analytic ansatz of Eq.\eqref{aps-2}, except at the resonance where $\lambda_+$ is a positive integer. 
Since $m^2>0$ for $\alpha>0$, $\lambda_+>0$  is always guaranteed  and the $s$-mode scalar always exists.

For a choice of $\lambda_+=1$, Eq.~\eqref{indicial} gives $m^2v_0=2$, i.e.
\begin{equation}
N_2=\frac{g''(\phi_h)P^2}{4r_+^4},
\label{resonance}
\end{equation}
which is exactly the exceptional case excluded when deriving constant scalar solution. 
The resonance is therefore real and it admits a closed form. 
Using $g''(\phi_h)=4\alpha$ for $g=1-\alpha\phi^2+\beta\phi^4$, Eq.~\eqref{resonance} reads $4\epsilon P^4/r_+^4=-\alpha P^2+r_+^2$, while Eq.~\eqref{loweq-1} gives $4\epsilon P^4/r_+^4=2(g_0P^2-r_+^2)$.
From these, one finds  $3r_+^2=P^2(\alpha+2g_0)$ and substituting  it back into Eq.~\eqref{loweq-1} determines the magnetic charge
\begin{equation}
r_+^2=\frac{P_{\rm res}^2\big(\alpha+2g_0\big)}{3},\qquad
P_{\rm res}^2=\frac{54\,\epsilon}{\big(\alpha+2g_0\big)^2\big(g_0-\alpha\big)}.
\label{Pres}
\end{equation}
The resonance occurs at a single magnetic charge $P_{\rm res}(\alpha,\beta,\epsilon)$ for each theory and only when $g_0>\alpha$.

On the other hand, the backreaction of Eq.~\eqref{frob} on the metric function  is fixed by Eq.~\eqref{neom1}. 
Balancing the $O(\rho^{2\lambda_+})$ term gives
\begin{equation}
N=N_2\rho^2-N_2 r_+\lambda_+ c^2\rho^{2\lambda_++1}+\cdots,
\label{Nback}
\end{equation}
so a degenerate horizon requires an inequality of  $\lambda_+\ge1/2$ ($m^2v_0\ge3/4$). 
Otherwise, the correction dominates and may induce negative  $N$.

Finally,  the local solution space at the degenerate horizon is strictly larger than the analytic ansatz dictated  by Sec.~\ref{sec:constant}  with an extra parameter of the single amplitude $c$.

\subsection{Extremal black holes with nonconstant scalar hair}\label{sec:running}

In what follows, we use the polynomial coupling of $g_{\rm poly}(\phi)=1-\alpha\phi^2+\beta\phi^4$. 
For the exponential coupling,  the condition of $g'(\phi)=0$ provides  the trivial root $\phi_h=0$ only.  We know that  no such an extremal solution exists, as shown in Sec.~\ref{sec:constant}.

Once the theory  with four  parameters $(P,\alpha,\beta,\epsilon)$ is given, none of the horizon data is free. 
The horizon scalar $\phi_h$ will be determined  by $g'(\phi_h)=0$, then one finds  the extremal horizon  $r_+$ from Eq.~\eqref{loweq-1}, the metric function  coefficient $N_2$ from Eq.~\eqref{loweq-3}, and the positive eigenvalue  $\lambda_+$ from Eq.~\eqref{lambdapm}.
The only remaining freedom is regarded as  the amplitude $c$ of the regular Frobenius mode~\eqref{frob} and it may be fixed by imposing the single condition of $\phi_\infty=0$.
The problem is therefore reduced to a one-parameter shooting problem with a one-dimensional target and for each $P$, and  it  may have an isolated solution.
In this case, it is worthy to note that the integration starts  from $\rho=\rho_0$, so that the neglected terms of Eqs.~\eqref{frob} and \eqref{Nback} lie below the integration tolerance.

\begin{figure*}[t!]
   \centering
    \includegraphics[width=0.32\textwidth]{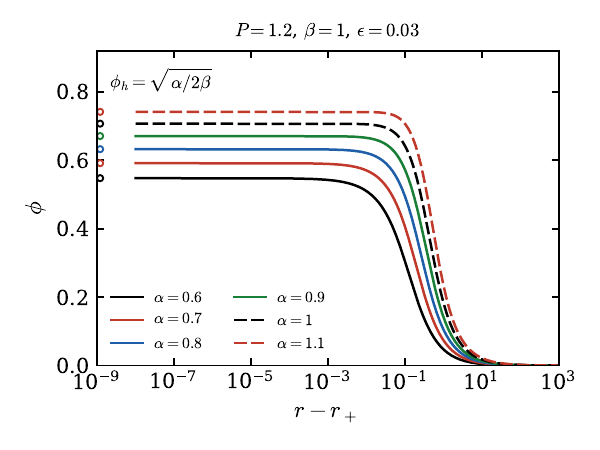}
    \includegraphics[width=0.32\textwidth]{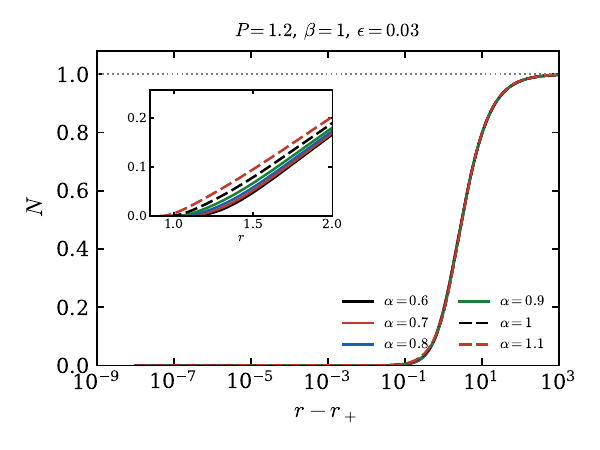}
    \includegraphics[width=0.32\textwidth]{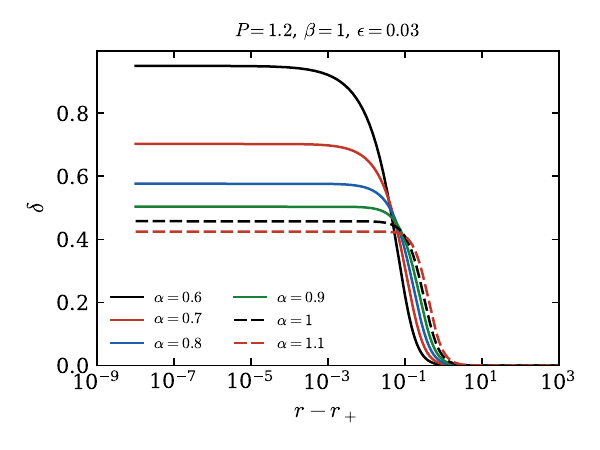}
\caption{Profiles of the scalarized extremal EEH black holes with nonconstant scalar hair, at fixed $P=1.2$, $\beta=1$, $\epsilon=0.03$ and six values of coupling constant $\alpha=0.6,0.7,0.8,0.9,1,1.1$, all with $\phi_\infty=0$.
(left) Scalar hair $\phi$ as function of $r-r_+$. (middle) Metric function $N$ as a function $r-r_+$. (right) Redshift function $\delta$ as a function of $r-r_+$. }
\label{fig:nonconst}
\end{figure*}

Searching for  $c$,  we find such solutions for each  $P$ above the threshold at which a degenerate horizon exists. 
Fig.~\ref{fig:nonconst} shows the profiles at fixed $P=1.2$ for six $\alpha$. 
As we can see in the middle panel of Fig.~\ref{fig:nonconst}, the metric function $N$ vanishes quadratically at $r_+$.  So one has $T_H=0$, while $N>0$ everywhere outside. 
The scalar hair remains at its attractor value of  $\phi_h=\sqrt{\alpha/2\beta}$ across more than six orders of magnitude in $(r-r_+)$, representing the AdS$_2$ throat. 
It turns over at $r-r_+\sim10^{-1}$ and decays to $\phi_\infty=0$ with scalar charge $Q_s>0$. 
The redshift function $\delta$ behaves in the same way, frozen at $\delta_0$ inside the throat and relaxing to zero outside. 
The quantity $\delta_0$ measures the total redshift accumulated across the throat and grows as $\alpha$ decreases towards the value at which $\lambda_+\to1/2$. 

The constant-scalar solution of Sec.~\ref{sec:constant} and the nonconstant-scalar solution are not two independent solutions, but they belong to a single local family described by the constant $c$. For $c=0$, one recovers the former solution.
As mentioned above, fixing the theory, $c$ is the only remaining parameter of the local solution \eqref{frob}-\eqref{Nback}, so that the asymptotic scalar $\phi_\infty(c)$ is a function of one variable at fixed $(P,\alpha,\beta,\epsilon)$. 
This map is monotonic with $\phi_\infty(0)=\phi_h$, so the constant-hair solution is located at the endpoint $c=0$.  On the other hand,  the physical solution of nonconstant hair is located at the successive $c_n$ satisfying  $\phi_\infty(c_n)=0$. Therefore, we could find infinite branches of  extremal scalarized black holes (See  Fig.~\ref{fig:qalpha} for the first three branches). In the following discussion, we focus  on the fundamental extremal branch with $n=0$.

\begin{figure*}[t!]
   \centering
    \includegraphics[width=0.45\textwidth]{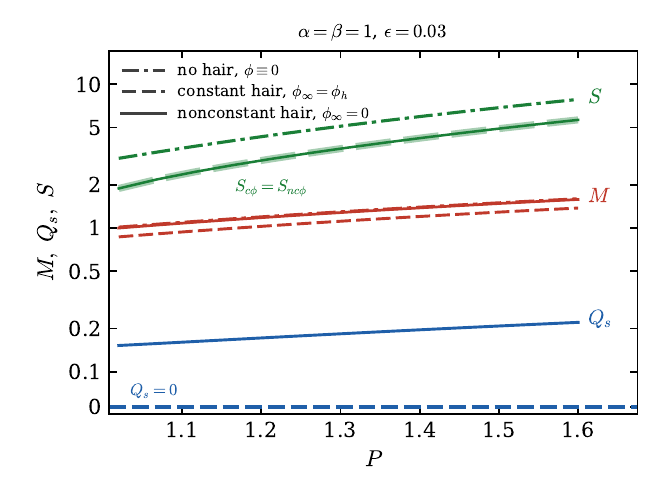}
    \includegraphics[width=0.45\textwidth]{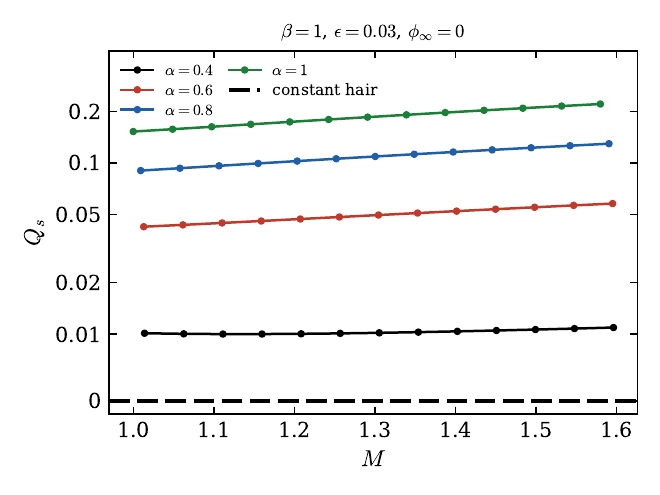}
\caption{Scalarized  extremal black holes at $\alpha=\beta=1$, $\epsilon=0.03$ for $n=0$. (left)  Mass ($M$), scalar charge ($Q_s$), and entropy ($S$) are functions of  $P\in[1.02,1.60]$  for three extremal families.  (right) Scalar  charge $(Q_s)$ as a  function of mass $M\in[1.0,1.6]$  for four different coupling constants $\alpha=0.4,\,0.6,\,0.8,\,1$.}
\label{fig:thermoext}
\end{figure*}

The left panel of Fig.~\ref{fig:thermoext}  shows  the thermodynamic quantities for scalarized extremal black holes with  constant-hair and nonconstant-hair over $P\in[1.02,1.60]$ at $\alpha=\beta=1$, together with the hairless extremal EEH black hole at the same charge discussed in Sec.~\ref{sec:theory}. 
Both scalarzied extremal families have $T_H=0$, and what is less obviously is that at the same $(P,\epsilon,\alpha,\beta)$, their entropies are also identical  because $(\phi_h, r_+)$ are shared by both endpoints as well as all intermediate members, regardless of the value that  the scalar approaches at infinity. 
This shows  the clearest manifestation of the attractor property~\cite{Ferrara:1995,FerraraKallosh:1996}.  Namely,  the horizon does not know the boundary condition  but it explains why the entropy \eqref{entropyC} obtained from the entropy function remains  unchanged when applying to the entire family and not only to the constant-hair scalar. 

Furthermore,  we observe  that the extremal hairy black hole is not entropically preferred. 
This is not in conflict with the non-extremal result of Sec.~\ref{sec:nonlinear}, where at fixed $(M,q)$ one found the desired case of $S>S_{\rm EEH}$. 
In the non-extremal case, the scalarized solution bifurcates from the hairless one at the same $(M,q)$, so the two are competing solutions of the same boundary value problem. 
However, once $P$ is located  at extremality with  fixed mass $M$, 
any two of the three families cannot  be placed at the same values $(M,P)$. 
Moreover, it is known that  the constant scalar hair member survives  with $\phi_\infty=\phi_h\ne0$ and it does not even belong to the same asymptotic sector. 
Fig.~\ref{fig:thermoext} therefore shows  the three extremal endpoints that  exist in this theory at a given charge, rather than a statement about which dynamical process would be selected. 
What that two solutions do not share is  the mass $M$ and the scalar charge $Q_s$, both of which are outputs obtained  from the shooting method. 
We note that the nonconstant-hair black hole is always heavier than constant-hair black hole and the former always carries scalar charge $Q_s>0$.

Finally, primary hair is diagnosed by whether the scalar charge can vary or not while the black hole hairs of mass and charge  together with action variables are fixed. 
In $(M,Q_s)$ plane of the right panel, the scalar charge increases as mass increases. As well, we notify from the left panel that the scalar charge  increases as charge $P$ increases. 
This indicates clearly that the scalar hair becomes secondary, in agreement with general expectations for extremal black holes~\cite{Zou:2019ays,Zou:2020rlv}.

\subsection{Connection between scalarzied  non-extremal  and extremal solutions}

\begin{figure*}[t!]
   \centering
    \includegraphics[width=0.45\textwidth]{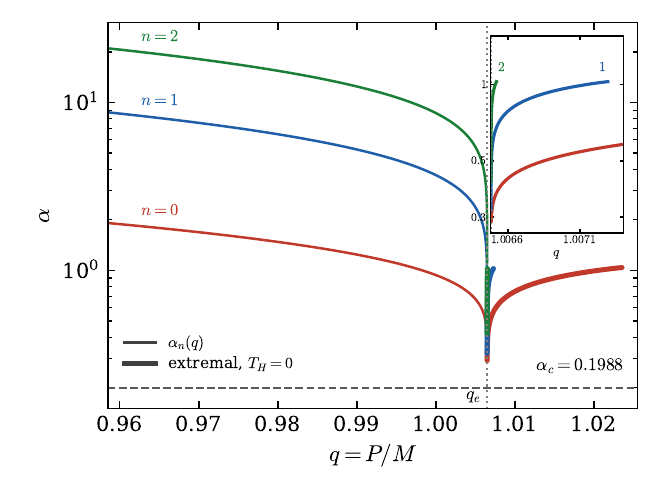}
    \includegraphics[width=0.45\textwidth]{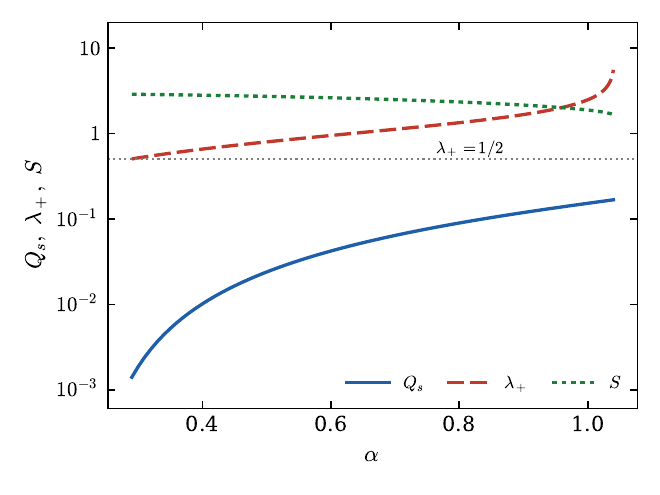}
\caption{ (left) Extremal branches for $\epsilon=0.03$ and $\beta=1$ in the $(q,\alpha)$ plane. Colored thin curves represent  the existence curves $\alpha_n(q)$ for  $n=0,1,2$ non-extremal branch, while colored thick curves denote the $n=0,1,2$ extremal branches for $T_H=0$.
Two boundaries of the scalarized solution domain meet only on the  dotted vertical line at extremal point of $q=q_e$. 
(right) Scalar charge ($Q_s$), throat exponent ($\lambda_+$) from Eq.\eqref{lambdapm}, and entropy ($S=\pi r_+^2$) are displayed  along the $n=0$ extremal branch.}
\label{fig:qalpha}
\end{figure*}

One important question remained is to ask whether the scalarized extremal solutions of Sec.~\ref{sec:running} accommodates the $T_H\to0$ endpoints of the scalarzied non-extremal branches constructed in Sec.~\ref{sec:nonlinear}. 
To handle this issue, one requires the same normalization. 
As noted after Eq.~\eqref{asyp-3}, the physical dimensionless parameters are $(q,\ \epsilon/M^2,\ \alpha,\ \beta)$, so the extremal solutions must be rescaled to $M=1$ at fixed $\epsilon=0.03$, exactly as did in Sec.~\ref{sec:cold}. 
Then, the results are displayed in Fig.~\ref{fig:qalpha}.
The construction is repeated for each extremal branch, so that the extremal diagram carries the same node numbers as the non-extremal side does have.

One can see from the left panel that the existence curves of the scalarzied non-extremal black holes approach $q_e$ from the left, while the existence curves of the scalarized extremal black holes emerges  from $q_e$ and extends toward $q> q_e$. 
In this case, the extremal point of the hairless black hole thus becomes a common accumulation point for the entire scalarization region, consistent with the linearized  analysis of Sec.~\ref{sec:accum}. This implies that the conclusion is confirmed at the nonlinear level. 
Moreover, as is shown Fig.~\ref{fig:thermo},  when the scalarized C-horizon black holes approach the extremal point $q_e=1.00648$, $T_H$  increases and it thereby deviates from the extremality. This means that the extremal hairy black holes are  not the endpoints of the non-extremal branches bifurcating from the hairless C-horizon black holes for $q<q_e$.
The extremal branches  do not extend over all couplings $q$. 
This is because the fundamental extremal  branch exists only in the  window of  $0.2864\ \le\ \alpha\ \le\ 1.042$ and the excited extremal  branches occupy slightly narrower windows nested inside it.

The lower bound of the branch can be determined from the expansion coefficients at the extremal horizon. 
According to Eq.~\eqref{Nback}, at the lower end where $\lambda_+\to1/2$, the scalar backreaction term of $\rho^{2\lambda_++1}$ is no longer small when comparing with $N_2\rho^2$. In this case,  the near-horizon (AdS$_2\times S^2$) throat  loses its self-consistency accordingly.  This threshold is fixed by the condition of  $m^2v_0=3/4$, leading to $\alpha_{\min}=0.2864$ for $M=1$. 
We note that it is close to, but strictly above, the critical branch $\alpha_c=0.19884$ derived in Sec.~\ref{sec:accum}, leaving a narrow strip of $\alpha_c<\alpha<\alpha_{\min}$ for $q\gtrsim q_e$ in which neither the non-extremal nor the extremal family exists. For $\alpha>\alpha_{\max}$, it is worth noting that  extremal hairy black holes no longer exist.
This is not because the scalar cannot survive, but that the charge required by the black hole exceeds the range allowed by the theory. 
At the point beyond $\alpha=\alpha_{\rm max}$,  both throat exponent  $\lambda_+$ and throat amplitude ($Q_s\sim c$) diverge, as shown in the right panel of Fig.~\ref{fig:qalpha}.

\section{Discussions }

A well-known scalarized extremal black hole was the BBMB black hole, described by mass $m$ and charge $Q$ in the Einstein-Maxwell-conformally coupled scalar theory~\cite{Bekenstein:1974sf}.
Its secondary scalar hair takes the form of $\phi(r)=\sqrt{m^2-Q^2}/(r-m)$ with black hole mass $m=\sqrt{Q^2+Q_s^2}$, but it blows up at the horizon.
Scalarized extremal black hole with primary scalar hair (independent scalar charge $Q_s$) seems not to exist because requiring extremality (asymptotic condition) turn primary into secondary~\cite{Zou:2019ays,Zou:2020rlv}.
The present work confirmed this result in the generic EEHS theory and, more importantly, identified the mechanism behind it.

In the present work, we have investigated extremal black holes with scalar hair in the generic EEHS theory with two scalar couplings to the Maxwell term. One is an exponential coupling [$g_{\rm exp}(\phi)=e^{-\alpha \phi^2}]$ and the other is its polynomial quartic form [$g_{\rm poly}=1-\alpha \phi^2+\beta \phi^4$].
We have carried out  scalarizations for the C-horizon EEHBH over the whole triple-horizon window of $q\in[q_*,q_e]$ with $q_*=0.9503$ and $q_e=1.00648$ when choosing a fixed EH parameter $\epsilon=0.03$, extending the earlier $q=1$ analysis~\cite{Guo:2026qib}. Particularly, we have chosen the C-horiozn because it includes the extremality only.  In this case, we found infinite branches of scalarized non-extemal  EEHBHs because of  adopting spontaneous scalarization.
Both for the linearized theory and  the fully nonlinear theory, scalarization becomes easier as the C-horizon cools as $q$ increases, as was indicated with decreasing threshold coupling constant ($\alpha_{\rm th}=\alpha_0$) from $2.05870$ at $q_*=0.951$ to $0.356479$ at $q_e=1.00648$.

Importantly, we have found  that the approach to extremality ($q\to q_e$) is universal. 
All branch curves of scalarized non-extremal black holes tend to accumulate at a single value of critical branch [$\alpha_c=r_e^4N_2/(4P^2)=0.19884$] dictated by the BF bound of the AdS$_2$ throat, following the logarithmic law Eq.\eqref{accum}. 
We note  that the corresponding RN value is $\alpha_c=0.25$, so the Euler-Heisenberg terms make extremal black holes easier to be scalarized. 
As was shown in Fig.~\ref{fig:accum}, we observed that many nonextremal branches of $\alpha_n(q)>\alpha_c$ accumulated at the extremality of $q=q_e$. This is regarded as a new and interesting observation. 

At the same time, the extremality is fatal to have a primary hair (an independent quantity), since the degenerate horizon enforces $g'(\phi_h)=0$ and it hence leads to $\phi_1=\delta_1=0$ and $Q_s=0$ in the analytic approach of scalarization. 
The corresponding scalarized extremal black hole with constant secondary hair existed only for the polynomial coupling $g_{\rm poly}$, and we recovered it exactly from the entropy function applied to the Bertotti-Robinson geometry with two paramters $(v_0,v_1)$ identified as the near-horizon data $(1/N_2,r_+^2)$ rather than treated as free parameters. Furthermore,  the secondary hair was  confirmed from the nonconstant scalar hair solution because its scalar charge $Q_s$ depends on the mass $M$ and charge $Q$ (see Fig.~\ref{fig:thermoext}). 

Concerning the connection between scalarized non-extremal and extremal black holes, we displayed the left panel of Fig.~\ref{fig:qalpha}.  
When mapping both families onto the $(q,\alpha)$ plane leads to  their transparent  relation.
The extremal branches  did not continue the non-extremal branches bifurcating from the hairless C-horizon for $q<q_e$, which heat up rather than cool down, but they occupy $q>q_e$.
Both boundaries terminated at $q=q_e$, so the extremal point of the hairless black hole represents the single accumulation point of the entire scalarized black hole solutions, at the linearized level through the critical branch of $\alpha_c$ as well as nonlinearly through the $T_H=0$ trajectories.

Consequently, it is not easy to find scalarized extremal black holes with primary scalar hair, and our analysis explained  the reason by exploring various scalarized extremal black hole solutions for EEHBHs. 
What that  the same picture holds for $g(\phi)=f(\phi)=1-\alpha\phi^2+\beta\phi^4$, in which the Euler-Heisenberg term is also coupled to the scalar,  remains an interesting question.

\vspace{1cm}

{\bf Acknowledgments}
 \vspace{1cm}

X.Y.C. is supported by National
Science Foundation of China (no: W2533026).
 Y.S.M. is supported by the National Research Foundation of Korea (NRF) grant funded by the Korea government(MSIT) (RS-2022-NR069013).
 
\end{document}